    \documentclass[12pt,psf,epsf]{article}
\usepackage[centertags]{amsmath}
\usepackage{amssymb}
\usepackage{graphicx}
\usepackage{epsfig}
\usepackage{ulem}
\usepackage[english]{babel}
\usepackage{array}
\usepackage{amsthm}
\usepackage{latexsym}
\usepackage[mathcal]{euscript}
\pdfoutput=1
\usepackage{epsfig}

 \usepackage{jheppub}
 \usepackage{hyperref}

\usepackage{subfigure}
\usepackage{psfrag}

\usepackage{epsfig}
\usepackage[latin1]{inputenc}
\usepackage{float}
\usepackage{graphicx}
\usepackage{cancel}
\usepackage{mathrsfs}
\usepackage{amssymb}
\usepackage{amsfonts}
\usepackage{amsmath}
\usepackage{slashed}
\usepackage{graphicx}
\usepackage{bm}
\usepackage{color}
\newcommand{\be}{\begin{equation}}
\newcommand{\ee}{\end{equation}}
\newcommand{\bea}{\begin{eqnarray}}
\newcommand{\eea}{\end{eqnarray}}
\newcommand{\bwt}{\begin{widetext}}
\newcommand{\ewt}{\end{widetext}}

\newcommand{\nn}{\nonumber}

\newcommand{\bi}{\begin{itemize}}
\newcommand{\ei}{\end{itemize}}

\usepackage{setspace}

\begin{document}

\title {Shaping contours of entanglement islands in BCFT}

\author{Dmitry S. Ageev}
\affiliation{Steklov Mathematical Institute, Russian Academy of Sciences, Gubkin str. 8, 119991
Moscow, Russia}

\emailAdd{ageev@mi-ras.ru}

\abstract{In this paper, we study the fine structure of entanglement in holographic two-dimensional boundary conformal field theories (BCFT) in terms of the spatially resolved quasilocal extension of entanglement entropy - entanglement contour. We find that the boundary induces discontinuities in the contour revealing hidden localization-delocalization patterns of entanglement degrees of freedom. Moreover, we observe the formation of ``islands'' where the entanglement contour vanishes identically implying that these regions do not contribute to the entanglement at all. We argue that these phenomena are the manifestation of entanglement islands discussed recently in the literature. We apply the entanglement contour proposal to the recently proposed BCFT black hole models reproducing the Page curve - moving mirror model and the pair of BCFT in the thermofield double state. From the viewpoint of entanglement contour, the Page curve also carries the imprint of strong delocalization caused by dynamical entanglement islands. }

\maketitle

\newpage
\section{Introduction}
The mechanism which is responsible for the emergence of the information about the black hole microstates and interior in the Hawking radiation is still not well understood in its full generality. Recently significant progress concerning these issues has been made with holographic duality as one of the central methods of investigation \cite{Penington:2019npb,Almheiri:2019psf,Almheiri:2019hni,Almheiri:2019qdq,Penington:2019kki,Chen:2020uac,Chen:2020hmv,Almheiri:2020cfm}. One of the central goals of the modern understanding of the information paradox, firewall phenomena \cite{Almheiri:2012rt}, fuzzball proposal \cite{Mathur:2005zp} and many other interesting developments in black hole physics is the deeper comprehension of the   Hawking radiation density matrix evolution  \cite{Almheiri:2020cfm}. The unitary black hole evaporation should be accompanied by the specific form of the entanglement entropy evolution, a so-called Page curve \cite{Page:1993df,Page:1993wv}. The entanglement entropy following the Page curve increases initially due to the thermal character of Hawking radiation and then decreases at late times, indicating the consistency with the unitary evolution.

Further understanding of this kind of evolution can be achieved by study of more fine-grained probes, for example decomposition of the entanglement entropy in some quasi-local quantity. In this paper we consider such probe describing the distribution of the entanglement inside the subsystem or in other words spatially resolved version of entanglement entropy, the kind of entanglement density fixed by some certain list of properties \cite{chen-vidal}. It is called the entanglement contour and has been considered recently in different contexts including out-of-equilibrium physics and holography \cite{Tonni:2017jom,Coser:2017dtb,Eisler:2019rnr,Wen:2018whg,Wen:2019ubu,Kudler-Flam:2019oru,Han:2019scu,Kudler-Flam:2019nhr,Ageev:2021iyw,MacCormack:2020auw,Han:2021ycp}.

Our main object to study in this paper from the viewpoint of entanglement contour is quite a wide class of quantum systems, namely boundary conformal field theories (BCFT) \cite{cardy}. Our goal here is two-fold. First, we would like to understand what new information in comparison to the entanglement entropy  contour function can tell us about the physical properties of general BCFT. Is it possible to reveal some hidden and more fine-grained features using this quantity? On the other hand, we would like to investigate how the entanglement contour resolves the Page curve in certain examples of two different black hole  BCFT models proposed recently \cite{Rozali:2019day,Sully:2020pza,Akal:2020twv,Akal:2021foz}. The Hawking radiation is believed to carry the imprint of the microscopic structure of black holes encoding the information about it. On the other hand the entanglement entropy subsequently monotonously increases and decreases providing us a quite simple picture of density matrix evolution at a first sight. So our second goal is to study what fine-grained features of Hawking radiation in holographic BCFT models could be revealed by entanglement contour.

One convenient and interesting class of models which is believed to mimic black hole properties and have similar energy  radiation is moving mirrors \cite{BD,mir1,mir2}. In this class of model, the boundary is not stationary and follows the prescribed trajectory leading to a non-linear response of quantum fields interacting with it. In \cite{Akal:2020twv,Akal:2021foz} it was shown that the Page curve arises as a result of the calculation in the holographic dual of such model (see for previous studies of moving mirrors in a holographic setup \cite{Bianchi:2014qua,Hotta:2015huj,Good:2016atu,Chen:2017lum}).

The second model we study is the pair of BCFT in the thermofield double (TFD BCFT model for short) proposed to be the model of a black hole which is in equilibrium with the Hawking radiation \cite{Rozali:2019day,Sully:2020pza}. While in the holographic mirror we situate the boundary on the mirror trajectory in the TFD-BCFT model we put it on a complex plane in such a way that path-integral on this geometry corresponds to a thermofield double.

We start with a holographic dual of static BCFT proposed in \cite{Fujita:2011fp,Takayanagi:2011zk} which is the simplest example to consider. It is known, that the entanglement entropy of the interval in (holographic) BCFT on half-line exhibits phase transition when its location is far away enough from the boundary. The origin of this phase transition is due to geodesic configurations with different topologies contributing to the entanglement entropy. It is well known that according to Hubeny-Rangamani-Ryu-Takayanagi prescription only the configuration with the minimal length contributes to the entanglement entropy \cite{Ryu:2006bv,Hubeny:2007xt}. This change of leading geodesic configuration is present also in the holographic mirror and TFD-BCFD models. In its essence, it provides the change of regime of entanglement entropy evolution from increasing to decreasing leading to the correct Page curve in the holographic mirror model. A similar transition shapes the Page curve in the models with braneworld and dilaton gravity \cite{Penington:2019npb,Almheiri:2019psf,Almheiri:2019qdq,Chen:2020uac}. The origin of the phase transition in these models is the presence of ``entanglement islands'' which contribute to the entanglement of the region under consideration but located somewhere else. In BCFT models the role of entanglement island is played by the  end-of-the-world brane which is responcible to the disconnected HRRT geodesics (see \cite{Rozali:2019day,Akal:2020twv}) and we also will use this notion in our context.

$\,$

We show that the presence of this phase transition leads to the unexpected structure of the entanglement contour in BCFT. Simple at first sight behavior of the entanglement entropy after the spatial resolution into the entanglement contour gets additional counterintuitive features. Let us briefly summarize our findings:
\begin{itemize}
    \item As expected the presence of boundary makes the entanglement contour strongly inhomogeneous. In the near-boundary zone, the entanglement contour for small enough values of boundary entropy vanishes identically. This implies that degrees of freedom in this zone do not contribute to the total entanglement of the region at all. This situation resembles the situation with the firewall paradox where the presence of entanglement near the horizon contradicts the entanglement across the horizon and between early/old Hawking radiation. One of the resolutions of this issue proposed in \cite{Yoshida:2019qqw} is that the presence of observer near the black hole leads to partial disentanglement in Hawking radiation modes resembling partial disentanglement observed here.
    
     \item In general, one can state that the manifestation of ``entanglement island''  discussed recently in literature \cite{Almheiri:2020cfm} can be observed as the ``islands`` even in the static entanglement contour. For example, these islands can split the region into two parts by the presence of a  disentangled zone.
    
    \item The Page curve for mirror and TFD BCFT models has a complicated structure after fine-graining by entanglement contour. One can observe a sophisticated
 the pattern of entanglement localization-delocalization consisting of many ``islands'' propagating during the black hole evaporation. 
   
\end{itemize}

This paper is organized as follows. We start with a brief review of the definition and basic properties of entanglement contour in Section 2. In Section 3 we study the entanglement contour in the static BCFT. Sections 4 and 5 are devoted to entanglement contour in the moving mirror model and the TFD BCFT models. We conclude with some remarks and future directions of research in Section 6.

\section{The entanglement contour}\label{sec:entc}
Roughly speaking the entanglement contour  of the subsystem  $A$ is the function $f_A(x)$  defined\footnote{For simplicity we assume that $A$ is the connected region.} as 
\be \label{eq:EC}
S(A)=\int_{x \in  A} f_A(x)dx,
\ee 
where $S(A)$ is the entanglement entropy of $A$ or in other  words it  is the density function of the entanglement entropy. Though the fundamental definition of the  entanglement contour is still missing one can restrict a possible class of functions using the following (incomplete) list of properties
\begin{itemize}
    \item Entanglement contour is a non-negative function:
\be     
    f_A(x)\geq 0 
\ee    
   
    \item The entanglement contour $f_A(x)$ inherits symmetries of  the reduced density matrix $\rho_A$.
    \item  The entanglement contour is invariant under local unitary transformations.
    \item Upper bound: if $ {\cal H}_{T} = {\cal H}_{B} \otimes {\cal H}_{\bar B}$ and ${\cal H}_X \subseteq {\cal H}_{B} $ then
    \begin{gather}
        f_{T}(x) \leq S(B).
    \end{gather}
\end{itemize}
In  \cite{Wen:2018whg} the proposal for a  contour function in terms of  the partial entanglement entropy has been given.  Given some one-dimensional system and the subsystems $A_1,A_2,A_3$ with $A=A_1\cup A_2\cup A_3$ the partial entanglement  entropy  $s_A(A_2)$ is  defined as 
\begin{gather} \label{eq:partS}
s_A(A_2)=\frac{1}{2}\Big(S(A_1 \cup A_2)+S(A_2 \cup A_3)-S(A_1)-S(A_3 )\Big),
\end{gather} 
quantifying the contribution of $A_2$ to the entanglement of the total system $A$. For the entanglement entropy of a single interval $(x_1,x_2)$ given by some function $S(x_1,x_2)$ in 1+1 dimensional theory with the spatial direction $x$  after taking the limit $A_2 \rightarrow 0$ it is straightforward to obtain the entanglement contour \cite{Kudler-Flam:2019oru} of this interval in the form
\be
f_A(x)=\frac{1}{2}\left(\frac{\partial S\left(x_{1}, x\right)}{\partial x}-\frac{\partial S\left(x, x_{2}\right)}{\partial x}\right).
\ee 
For convenience let us briefly list some simple well-known examples \cite{Wen:2018whg,Kudler-Flam:2019oru} of entanglement contour in two-dimensional conformal field theory. The entanglement entropy and related contour of the ground state is given by
\be
S(x_1,x_2)=\frac{c}{3}\log\left(\frac{x_2-x_1}{\varepsilon}\right),\,\,\,\,f_A(x)=\frac{c \left(x_2-x_1\right)}{6 \left(x-x_1\right)
   \left(x_2-x\right)},
\ee 
while the generalization on the finite temperature case has the form 
\begin{gather}
    S(x_1,x_2)=\frac{c}{3}\log\left(\frac{ \sinh (\pi T (x_2-x_1))}{\pi T \varepsilon}\right),\\f_A(x)=\frac{\pi  c T}{6}  \left(\coth \left(\pi  T
   \left(x-x_1\right)\right)+\coth \left(\pi  T
   \left(x_2-x\right)\right)\right).
\end{gather}
The contour functions diverge near the endpoints and takes its minimum in the center of entangling interval. The divergent term comes from the entanglement between infinite number degrees of freedom across the junction of interval and its complement.
\section{Static BCFT}
\begin{figure}[h!]
\centering 
\includegraphics[width=7.cm]{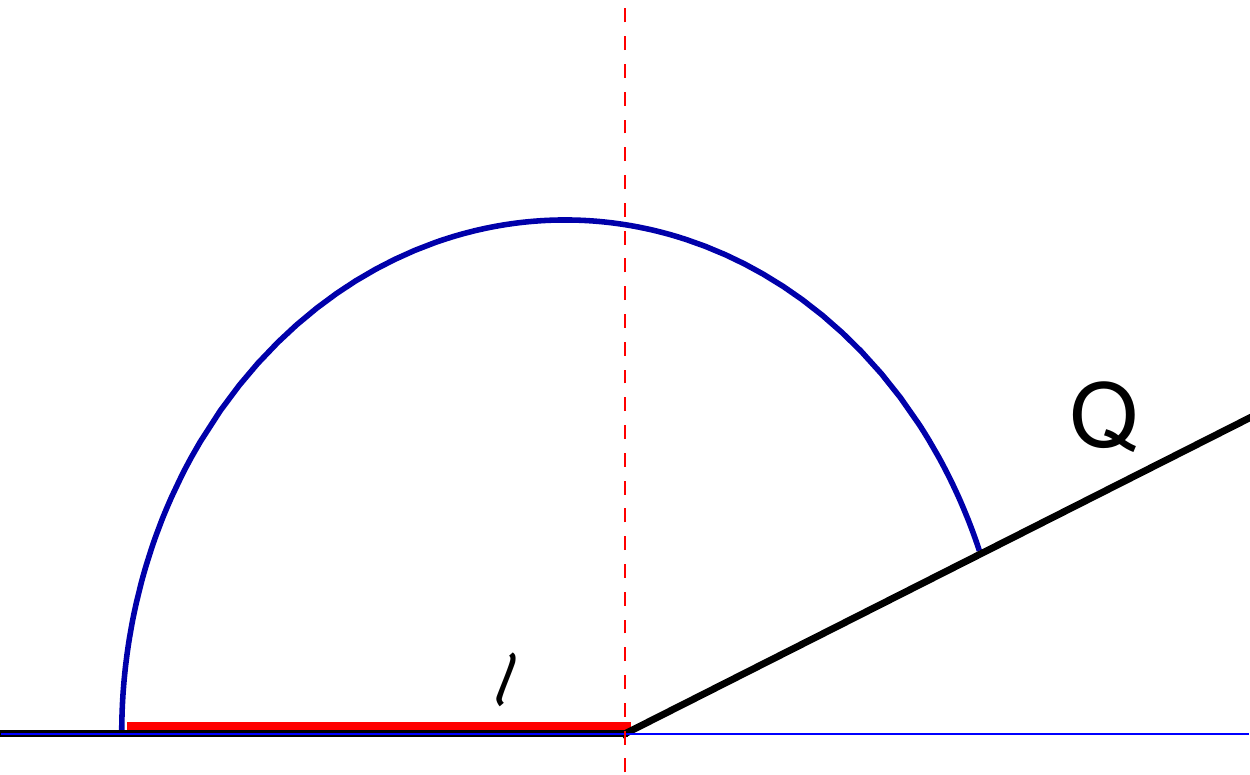}
 \caption{The HRRT surface corresponding to the entanglement entropy of the interval including the boundary (red solid line). The blue curve is the HRRT surface and $Q$ is the ETW brane.}
 \label{fig:adsrt1}
\end{figure}

\begin{figure}[h!]
\centering 
\includegraphics[width=6.5cm]{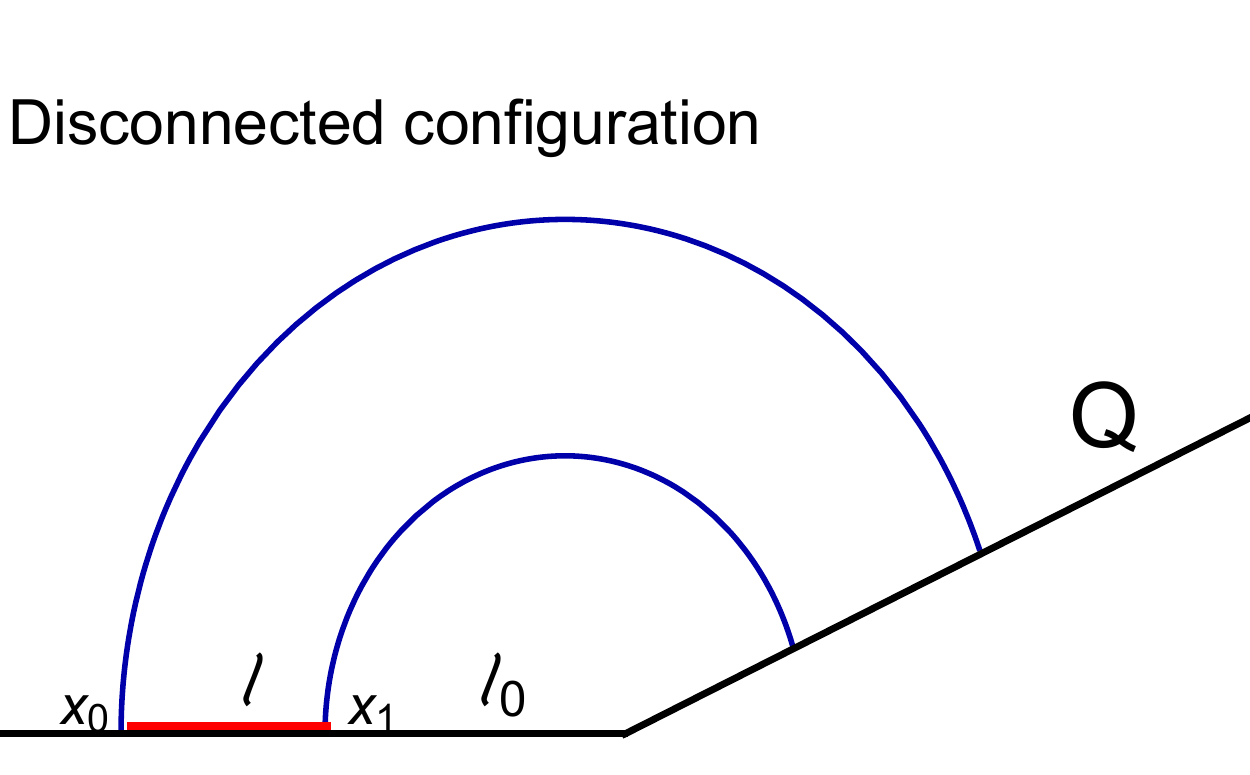}
\includegraphics[width=6.5cm]{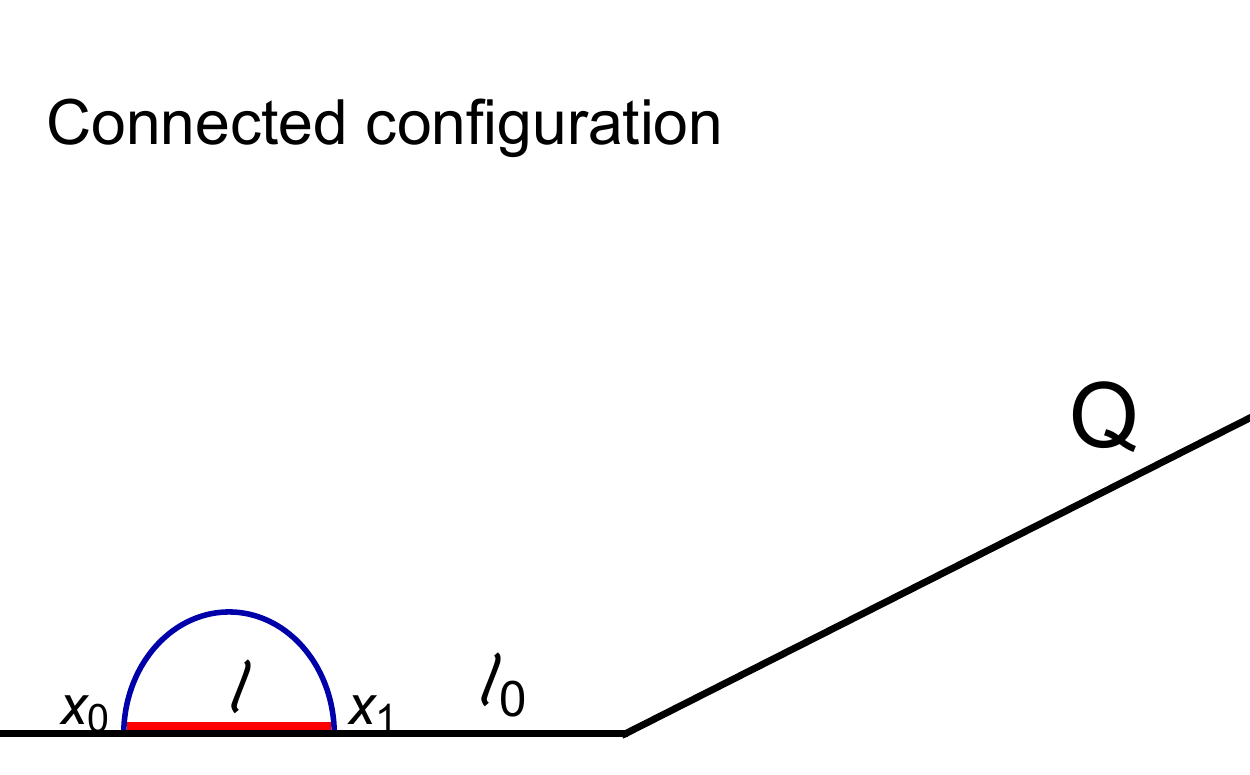}
 \caption{Different geodesic configurations contributing to the entanglement entropy of the interval $A$ of the length $\ell$ (red) placed on the distance $\ell_0$ from the boundary. Here $Q$ is the ETW brane given by \eqref{eq:poincETW} with the parameters $T=0.4, L=1$ fixed, and blue curves are RT surfaces spanned on the boundary of entangling interval $A$. On the left plot, we present disconnected geodesic configuration, and on the right plot the connected one.}
 \label{fig:adsrt}
\end{figure}

\begin{figure}[h!]
\centering
\includegraphics[width=7.7cm]{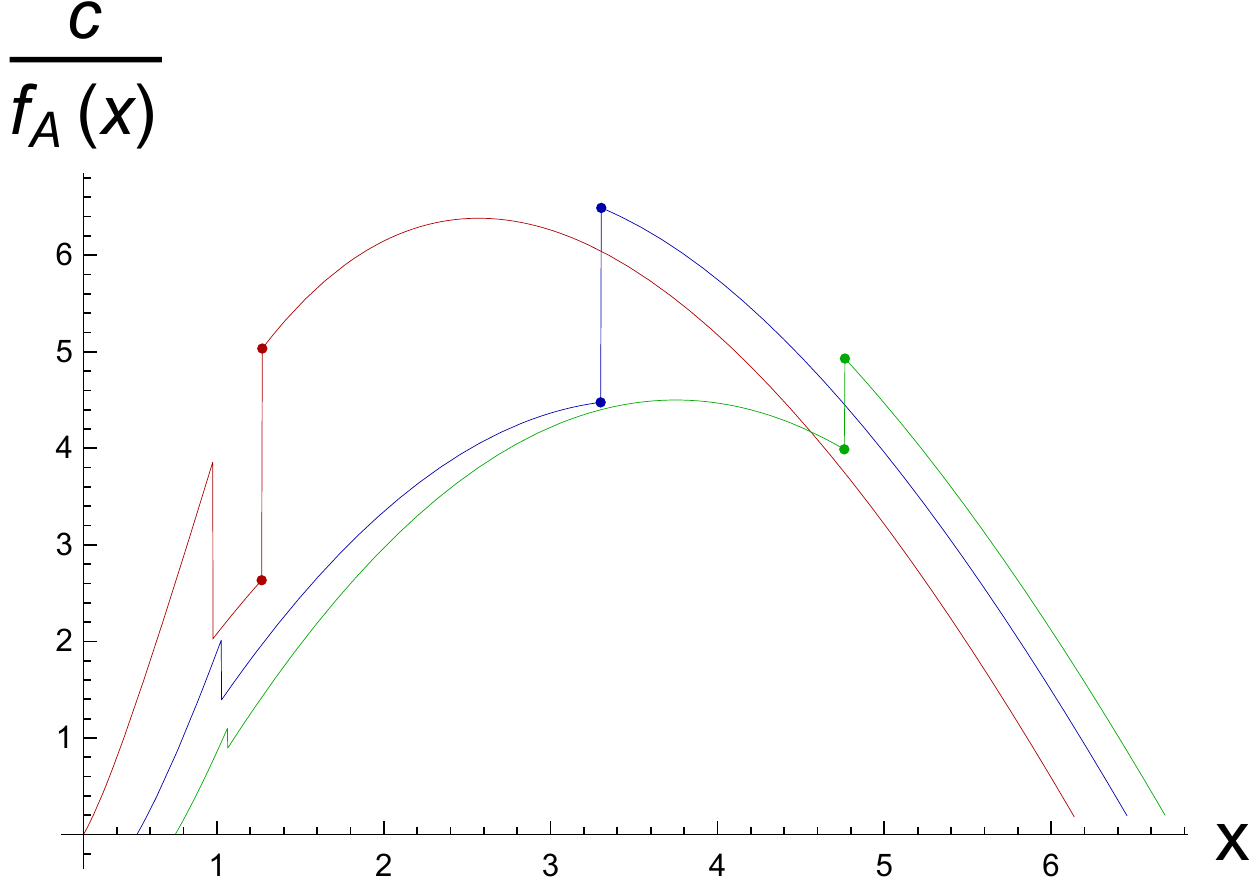}
\includegraphics[width=7.7cm]{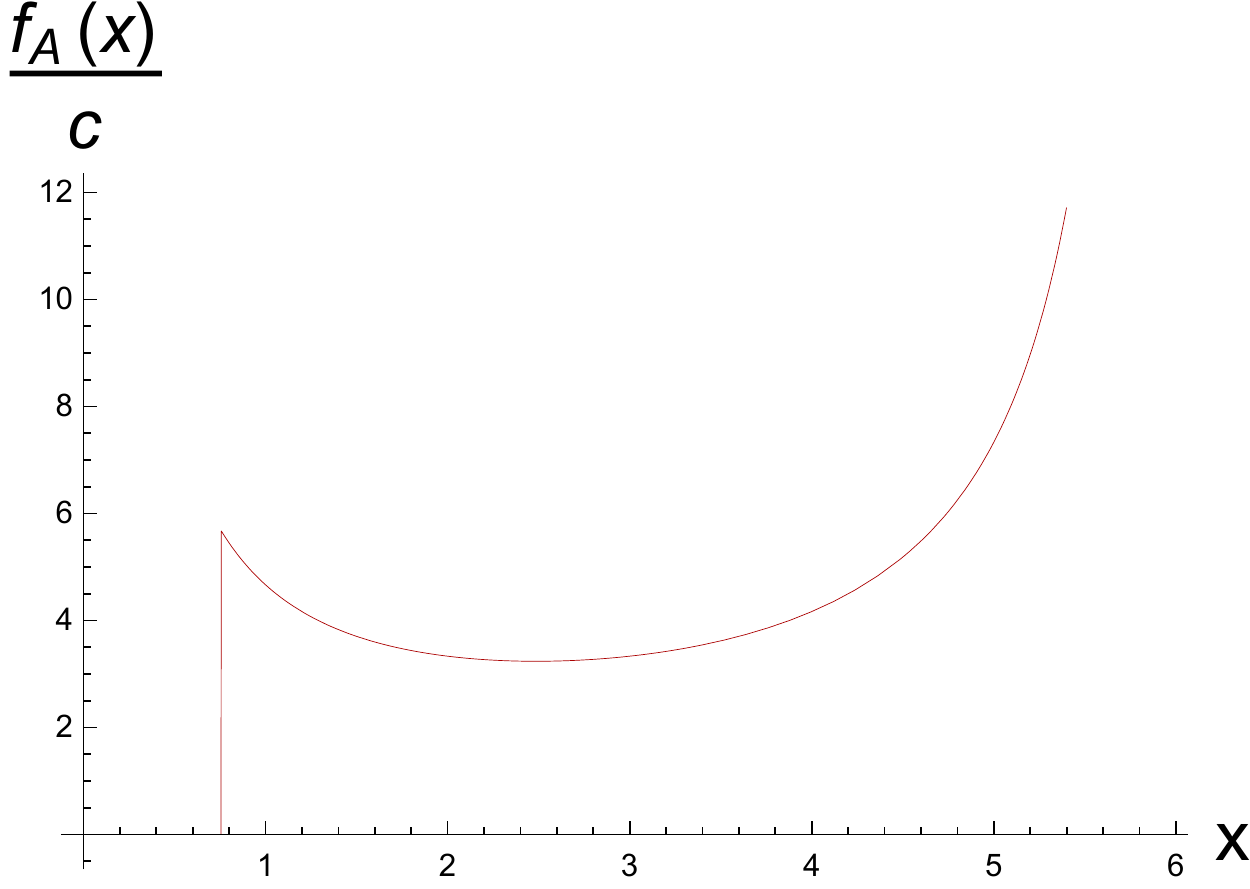}
 \caption{\emph{Left plot}: the inverse entanglement contour for the intervals of length $\ell=6$ and for different distances to the boundary $\ell_0=0.16$ (red curve), $\ell_0=0.52$ (blue curve) and $\ell_0=0.75$ (green curve). The values of central charge and boundary entropy are fixed to be $c=7$ and and $S_{bnd}/c=0.013$. 
  \emph{Right plot}: the entanglement contour for the interval including the boundary (i.e. $\ell_0=0$), with $\ell=6$ and  $S_{bnd}/c=0.035$).  }
 \label{fig:adsintEC}
\end{figure}

The construction of  BCFT holographic dual is based on the insertion of the ETW brane $Q$  consistent with the prescribed boundary conditions in the gravitational background \cite{Takayanagi:2011zk,Fujita:2011fp}.
The gravitational action including additional boundary terms due to ETW brane   has the form
\be \label{eq:GR}
I=\frac{1}{16 \pi G_{N}} \int_{N} \sqrt{-g}(R-2 \Lambda)+\frac{1}{8 \pi G_{N}} \int_{Q} \sqrt{-h}(K-T_{br}) ,
\ee 
 where $K$ is the trace of extrinsic curvature $K_{ab}$ and the constant $T_{br}$ is interpreted as the tension of the brane $Q$.
The equation of motion for $Q$ with the induced metric $h_{ab}$ has the form
\be \label{eq:Kab}
K_{a b}=(K-T_{br}) h_{a b},
\ee 
or after taking the trace
\be 
K=\frac{d}{d-1} T_{br}.
\ee 
The  dual of finite temperature 2d BCFT is known to be a one-sided BTZ black hole with the metric
\be \label{eq:BTZ}
ds^2=\frac{L^2}{z^2}\left(-f(z)dt^2+\frac{dz^2}{f(z)}+dx^2\right), \,\,\,\, f(z)=1-z^2/z_h^2,
\ee 
and the ETW brane in the bulk given by 
\be \label{eq:BTZbrane}
x(z)=\pm z_{H} \cdot \operatorname{arcsinh}\left(\lambda z\right),\,\,\,\,
\lambda=\frac{L T_{br} }{z_{H} \sqrt{1-L^{2} T_{br}^{2}}}.
\ee 
For $z_h\rightarrow \infty$ the metric reduces to the Poincare patch of three-dimensional AdS
\be \label{eq:poinc}
ds^2=\frac{L^2}{z^2}\left(-dt^2+dz^2+dx^2\right), 
\ee 
and the brane is just a plane given by
\be \label{eq:poincETW}
x=\pm \lambda z.
\ee 
In the  holographic duality the entanglement entropy $S_A$ of the subsystem $A$ in CFT is given by the  HRRT  formula \cite{Ryu:2006bv,Hubeny:2007xt} relating $S_A$ and extremal codimension-2 surface $\gamma_{A}$ (i.e. a geodesic for three-dimensional gravity) spanned on the boundary of $A$
\be 
S_{A}=\frac{\operatorname{Area}\left(\gamma_{A}\right)}{4 G_{N}}.
\ee 
In BCFT there are additional configurations of the HRRT surfaces connecting the boundary and the ETW brane. For the simplest case when the subsystem is the interval including the boundary $x\in (0,\ell)$  there is only one geodesic connecting $x=\ell$  and the brane (see Fig.\ref{fig:adsrt1}). The entanglement entropy, in this case, has the form
\be S(\ell)=\frac{c}{6} \log \left(\frac{2 \ell}{\epsilon}\right)+ \log g_{b},
\ee 
where the constant $g_b$ is related to the brane tension $T_{br}$ as
\be 
S_{bnd}=\log g_{b}=\frac{c}{6} \operatorname{arctanh}(L T_{br}).
\ee 
and this constant defining the boundary entropy $S_{bnd}$ is given by the part of HRRT surface with $x>0$ (i.e. on the right-hand side from the red dashed line in Fig.\ref{fig:adsrt1}).
As we will see further the presence of these geodesics introduces the nontrivial spatial structure of the entanglement contour even for equilibrium setup. As a warm-up consider the entanglement entropy of a single interval in the dual of \eqref{eq:poinc} (not necessary including the boundary). There are two competing configurations of geodesics which are presented in Fig.\ref{fig:adsrt}.  
We have to minimize over all possible configurations and roughly speaking the HRRT surface with disconnected topology (i.e. with one of the endpoints fixed on the ETW brane) dominates for small  $\ell_0$, while for large $\ell_0$  the connected topology contributes. Thus far away enough from the boundary we will observe a kind of ``phase transition'' in the entanglement entropy. The phase transition between different HRRT surfaces in its essence is the central point in the modern explanation of Page curve behavior in the black hole physics \cite{Penington:2019npb,Almheiri:2019psf,Almheiri:2019qdq,Chen:2020uac}.  For our further considerations, we need an explicit description of this transition. The entanglement entropy  $S(x_1,x_2)$ of the interval $[x_1,x_2]$ is given by the composition of disconnected and connected phase
\begin{gather} \label{eq:EE-x1x2}
S(x_1,x_2)=S^{cn}(x_1,x_2)\theta({\cal C}(x_1,x_2)-c)+S^{dc}(x_1,x_2)\theta(c-{\cal C}(x_1,x_2)) \quad \\ 
S^{\operatorname{cn}}(x_1,x_2)=\frac{c}{3} \log \left(\frac{x_{2}-x_{1}}{\epsilon}\right),  \quad
S^{\mathrm{dc}}(x_1,x_2)=\frac{c}{6} \log \left(\frac{4 x_{1}x_{2}}{\epsilon^2}\right)+2 \log g_{b},
\end{gather}
where $\theta(x)$ is the Heaviside step function and the function ${\cal C}(x_1,x_2)$  takes into account which topology contributes to the entanglement entropy for a fixed $x_{1,2}$. It is defined as the solution of $S^{\operatorname{cn}}(x_1,x_2)=S^{\operatorname{dc}}(x_1,x_2)$ and has the explicit form
\be 
{\cal C}(x_1,x_2)=-\frac{12 \log \left(g_b\right)}{\log \left(\frac{4 x_1
   x_2}{\left(x_2-x_1\right){}^2}\right)}.
\ee 
In what follows we also use the notation $S_{bnd}=\log g_b$. The entanglement contour of the entanglement entropy defined by 
\eqref{eq:EE-x1x2} has the form
\be \nn
f_{A}(x)=\frac{1}{6} c \left(\frac{2 \theta   \left(c-c\left(x,x_2\right)\right)}{x_2-x}+\frac{\theta
   \left(c\left(x_1,x\right)-c\right)-\theta
   \left(c\left(x,x_2\right)-c\right)}{x}+\frac{2 \theta
   \left(c-c\left(x_1,x\right)\right)}{x-x_1}\right),
\ee 
and one can see the presence of discontinuities in the entanglement contour from this formula.  We present the structure of the (inverse) entanglement contour of the interval  in Fig.\ref{fig:adsintEC}. 
Naively one can expect one discontinuity in the entanglement contour corresponding to the phase transition in the entanglement entropy. However, in contrast with these expectations one can observe, that there are two discontinuities in the bulk of the interval $A$ which strongly depends on the size of the interval $\ell$, distance to the boundary $\ell_0$ and boundary entropy $S_{bnd}$ revealing strong non-local effects in the spatial entanglement pattern.

\begin{figure}[h!]
\centering
\includegraphics[width=8.5 cm]{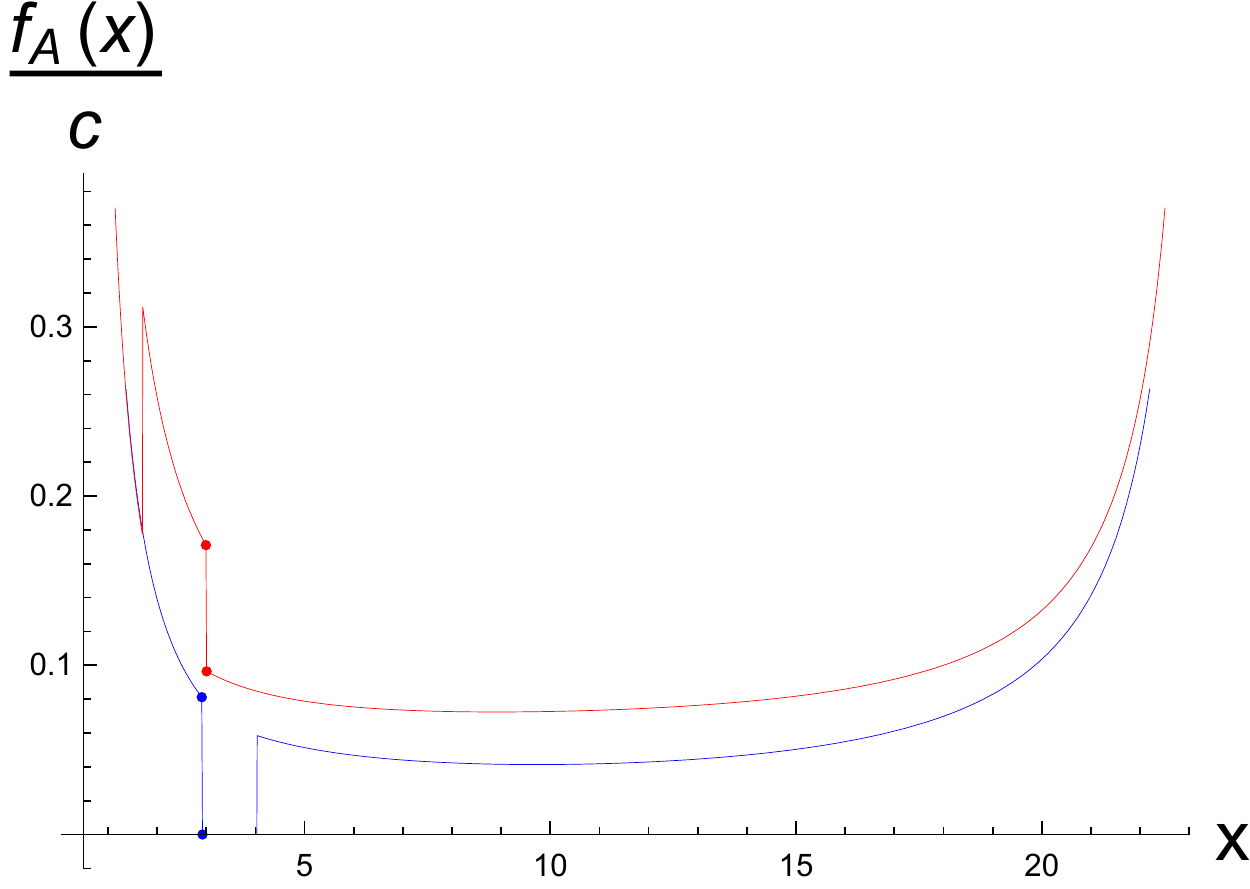}
 \caption{The normalized entanglement contour for the interval $x\in (0.5,23)$ and $S_{bnd}=0$. Blue curve corresponds to zero temperature entanglement contour, red curve to the temperature $T=0.03$. }
 \label{fig:ec1}
\end{figure}

What is even more curious one check that the entanglement contour in the limit  $\ell_0\rightarrow 0$ (i.e when the interval contains the boundary)  vanishes up to some point defined by the function ${\cal C}(x_1,x_2)$. The size of this zone grows with the decrease of the boundary entropy (see a right plot of Fig.\ref{fig:adsintEC}). This shows that the near-boundary zone does not contribute at all to the total entanglement and in some sense, it is disentangled with the rest of the system. The inclusion of the boundary in the entangling interval is not a necessary condition for vanishing the entanglement contour. If the region is situated close enough to the boundary the ``disentangled island'' peculiarly appears in the bulk of the region separating it into two parts (see blue curve in Fig.\ref{fig:ec1}).

$\,$

Let us discuss this result before turning to the non-equilibrium situation in the next sections:
\begin{itemize}
    \item Remind that here we follow the lines of \cite{Rozali:2019day,Sully:2020pza} that essentially the entanglement in TFD BCFT model (which is argued to be a model of a two-sided 2D black hole coupled to auxiliary
radiation) and ordinary BCFT on half-line are essentially driven by the same mechanism.  The observer who has the access to the region containing the boundary degrees of freedom will meet some strange region with the degrees of freedom disentangled from the rest of the system. The same should also take place in the other models of black holes. 

\item A similar situation appears in the firewall paradox. One of the main points in this paradox is that any  Hawking quantum radiated by an old black hole is entangled with the early radiation, which implies that it cannot be entangled with the modes behind the horizon. One of the possible resolutions is that the presence of the observer disentangles in part some degrees of freedom present in early and old Hawking radiation \cite{Yoshida:2019qqw}. In BCFT this manifests itself in spatially resolved entanglement --  close enough to the boundary, some region does not contribute to the entanglement in the region accessible to observer. What strengthens this line of reasoning is that this region is present only for the region with small enough boundary entropy. This is also in line with the black hole analogy because the old enough black hole corresponds to smaller entropy. 

\item  Without access to the boundary, the observer will find out the special ``island'' where the entanglement is especially weak. We state that this zone, in general, is the explicit manifestation of the entanglement island induced by the ETW brane (see \cite{Akal:2021foz} for a general discussion on the relation between ETW brane and entanglement island construction in different models). If the interval is large enough the entanglement contour may vanish separating the interval by the ``disentangled'' zone.

\item  The size of the  special zones  described above depends strongly on the boundary entropy. If the value of boundary entropy is larger than some critical value the entanglement contour becomes smooth again.
\end{itemize}
Finally, consider the entanglement contour in two-dimensional CFT on the half-line at finite temperature $T$. For simplicity consider the zero-tension ETW brane given by \eqref{eq:BTZbrane} corresponding to $S_{bnd}=0$. The dual background, in this case, is just the BTZ black hole being cut along the ETW brane\footnote{Notice, that if we consider not the half-line but the interval BCFT there is a kind of Hawking-Page transition between thermal $AdS_3$ and BTZ black hole for some certain $T$ and $T$.} given by equation $x=0$.  For this dual background the entanglement entropy of the system in different phases is given by
\begin{gather}
S^{\operatorname{cn}}(x_1,x_2)=\frac{c}{3} \log \left(\frac{\sinh\left(\pi T (x_{2}-x_{1})\right)}{\varepsilon \pi T}\right), \\ 
S^{\mathrm{dc}}(x_1,x_2)=\frac{c}{6} \log \left(\frac{\sinh\left(2 x_{1} \pi T\right)}{\varepsilon \pi T}\right)+\log \left(\frac{\sinh\left(2 x_{2} \pi T\right)}{\varepsilon \pi T}\right),
\end{gather}
where $T$ is the temperature of 2d CFT state. From Fig.\ref{fig:ec1} one can see that the small increase of the temperature strongly perturbs the state and the ``disentangled'' zone is replaced by the region with considerably amplified entanglement contour located near the boundary.
\section{Moving mirror}
\begin{figure}[h!]
\centering 
\includegraphics[width=7.5cm]{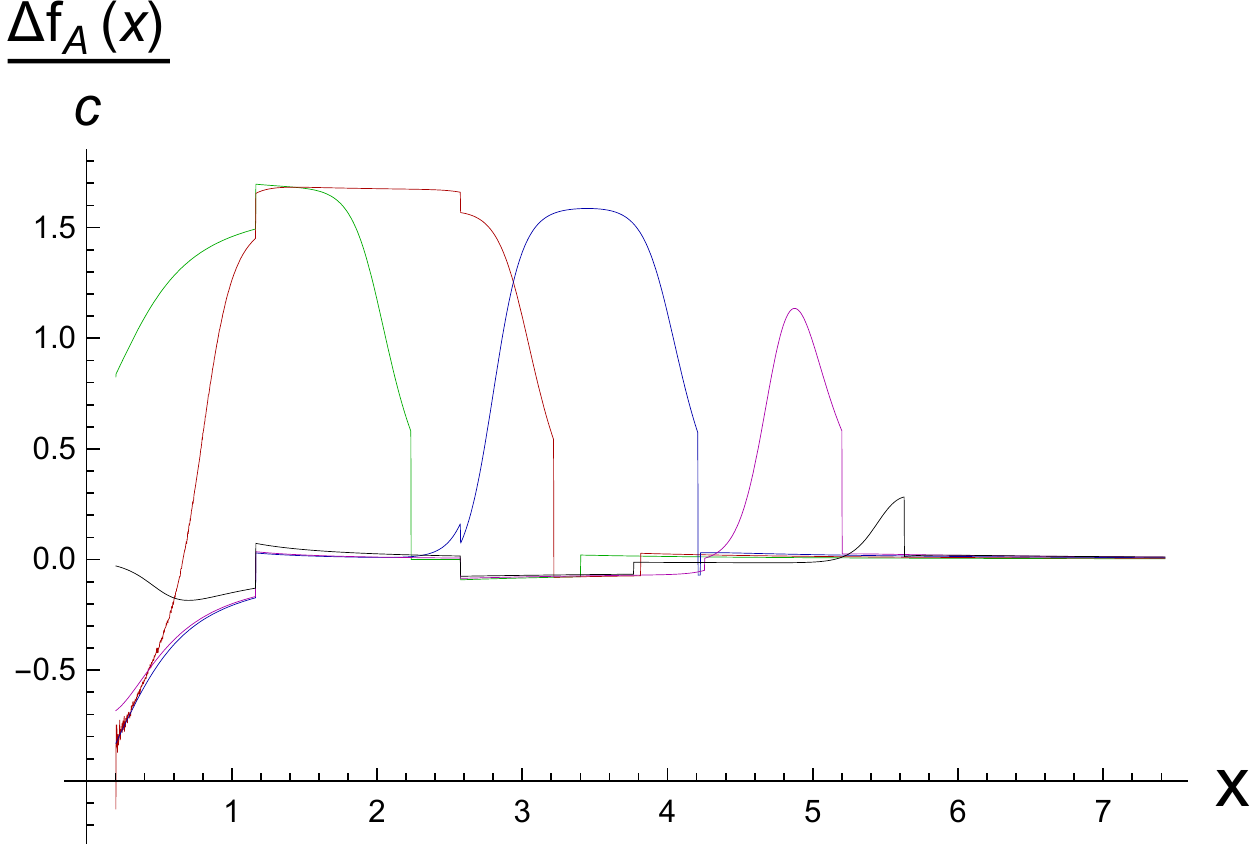}
\includegraphics[width=7.5cm]{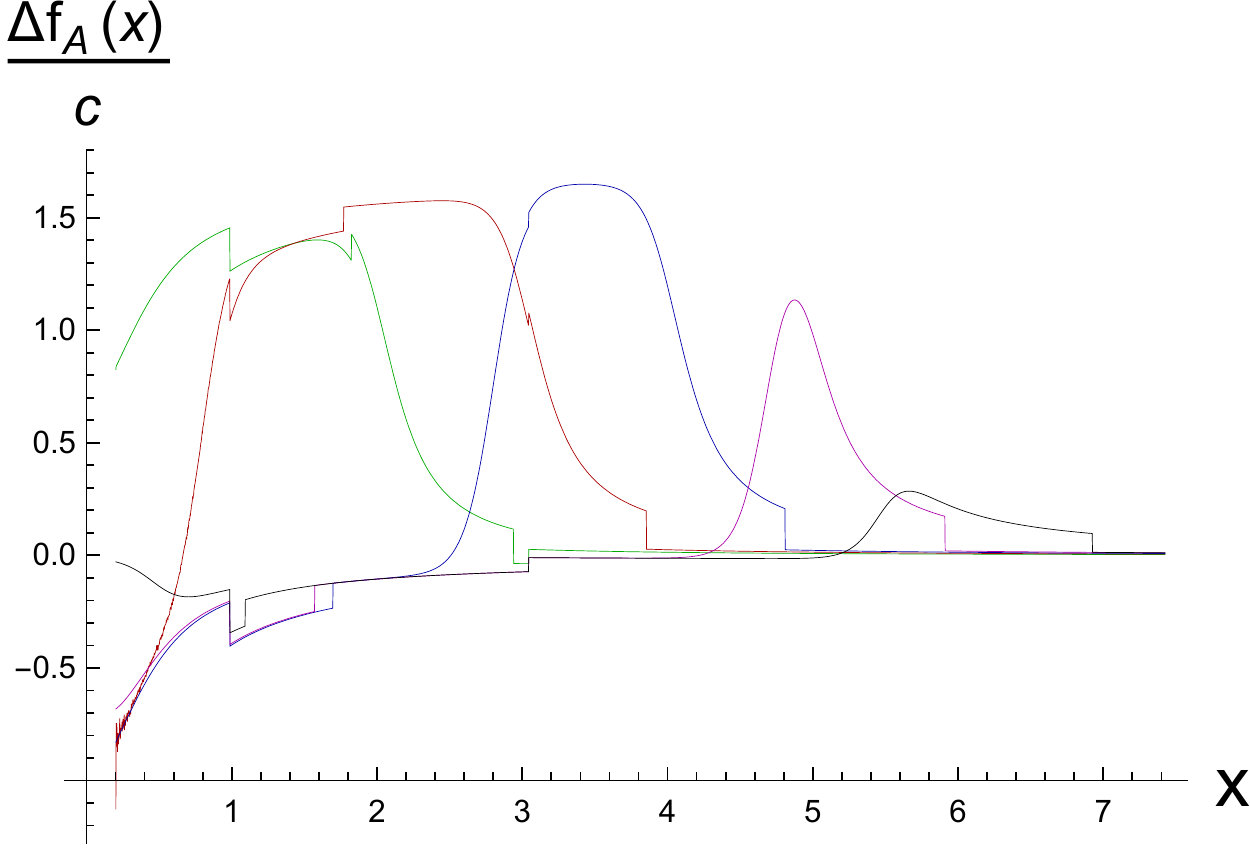}
 \caption{The entanglement contour of radiation in moving mirror model minus entanglement contour of the ground state at different time moments $t$. Green curves correspond to $t=1$, red $t=1.5$, blue $t=2$, magenta $t=2.5$ and black to $t=3$. For left plot boundary entropy $S_{bnd}=2\log g_b=0$, while $S_{bnd}=0.1$  for right plot. The entangling interval is chosen to be $x\in(0.2,15)$. }
 \label{fig:mirinf}
\end{figure}

A simple, solvable, and convenient to study model that mimics Hawking radiation in boundary field theory is a so-called ``moving mirror'' \cite{mir1}.  In this model the spatial location of the boundary is time-dependent and this dependence is chosen to capture particular black hole properties. In \cite{Akal:2020twv} the holographic realization of the moving mirror model has been proposed.  Starting with the fixed  trajectory $x=X_0(t)$ of the mirror and introducing lightcone coordinates
\be 
u=t-x,\,\,\,\,v=t+x
\ee 
one can show that the conformal mapping
\be  \label{eq:ccbnd}
\tilde{u}=p(u), \quad \tilde{v}=v, \quad t+X_0(t)=p(t-X_0(t)),
\ee 
maps the mirror with the trajectory $x=X_0(t)$ to the static one $\tilde{u}=\tilde{v}$. It is straightforward to find the gravity dual of this construction extending the coordinate transformation \eqref{eq:ccbnd} into the bulk as
\be \label{eq:ccbulk}
U=p(u), \quad V=v+\frac{p^{\prime \prime}(u)}{2 p^{\prime}(u)} z^{2}, \quad Z=z \sqrt{p^{\prime}(u)},
\ee 
which reduces to \eqref{eq:ccbnd} for $z\rightarrow 0$. This mapping relates the  Poincare $\mathrm{AdS}_{3}$ with the ETW brane given by \eqref{eq:poincETW} in lightcone coordinates $U$ and $V$ 
\be \label{eq:adslightcone}
d s^{2}=\frac{d Z^{2}-d U d V}{Z^{2}},
\ee 
and the background with the boundary along the prescribed mirror trajectory $X_0(t)$ with the ETW brane hanging from $X_0(t)$ into the bulk. The choice of the function $p(u)$ reproducing some certain properties of black hole evaporation is given  by 
\be \label{eq:pubh}
p(u)=-\beta \log \left(1+e^{-\frac{u}{\beta}}\right)+\beta \log \left(1+e^{\frac{u-u_{0}}{\beta}}\right).
\ee 
 The geodesics with different topologies contributing to the entanglement entropy can be obtained by conformal mapping. The entanglement entropy corresponding to these geodesics is explicitly given by
\begin{gather}\label{eq:EEmirr}
S_{A}^{\mathrm{ds}}= \frac{c}{6} \log \frac{(t+x_{0}-p\left(t-x_{0}\right))(t+x_{1}-p\left(t-x_{1}\right))}{\epsilon^2 \sqrt{p^{\prime}\left(t-x_{0}\right)p^{\prime}\left(t-x_{1}\right)}}+2 \log g_b, \\
S_{A}^{\mathrm{cn}}= \frac{c}{6} \log \frac{\left(x_{1}-x_{0}\right)\left[p\left(t-x_{0}\right)-p\left(t-x_{1}\right)\right]}{\epsilon^{2} \sqrt{p^{\prime}\left(t-x_{0}\right) p^{\prime}\left(t-x_{1}\right)}}.
\end{gather}
In \cite{Akal:2020twv} it was shown that for the interval moving parallel to the boundary (i.e. interval $(x_0+X_0(t),x_1+X_0(t))$) and choice $p(u)$ in the form \eqref{eq:pubh} the entanglement entropy follows the Page curve if $x_1 \rightarrow \infty$.  For $t\rightarrow \pm \infty$ we have the stationary boundary and at the intermediate time $t\approx 0$ we have $X_0(t) \approx -t $.  If  $x_1$ is finite the ``Page curve'' is repeated after some time capturing the entanglement corresponding to the particles reflected from the boundary and crossing the interval. We present the evolution of the normalized difference between the entanglement contour of radiation due to the mirror and the entanglement contour of the BCFT ground state in Fig.\ref{fig:mirinf}. We see that the simple at first sight evolution of entanglement which consists only of the linear growth and decrease has  complicated fine-grained structure which can be summarized as follows
\begin{itemize}
    \item One can observe at least four discontinuities in the entanglement which evolve and spread over the entanglement contour. For example (see the black curve in Fig.\ref{fig:mirinf})  the influence of the boundary during the evolution is present in the center of entangling interval at intermediate times.
    \item As usual for large boundary entropy, one can observe more smooth behavior of evolving entropy.
\end{itemize}

\section{Pair of BCFT in the thermofield double state}
\begin{figure}[h!]
\centering 
\includegraphics[width=7.5cm]{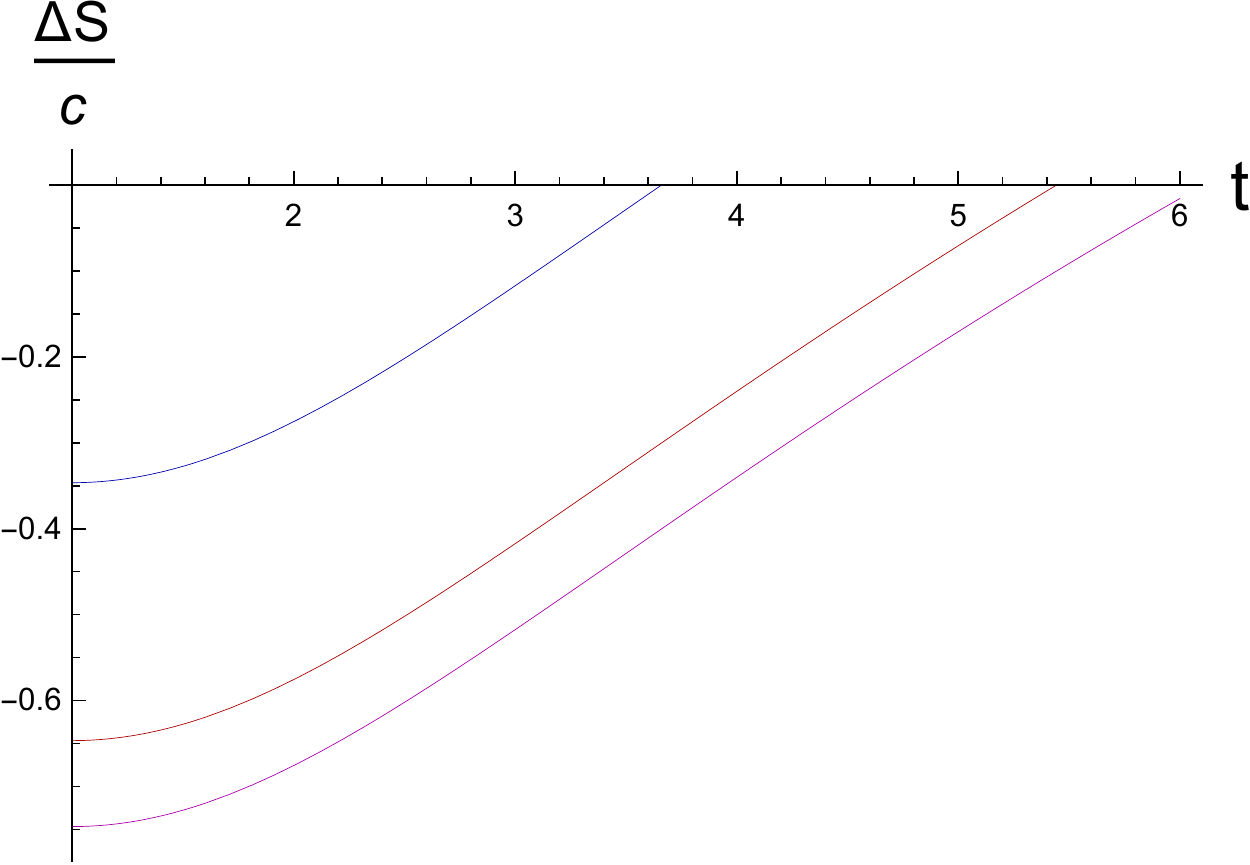}
\includegraphics[width=7.5cm]{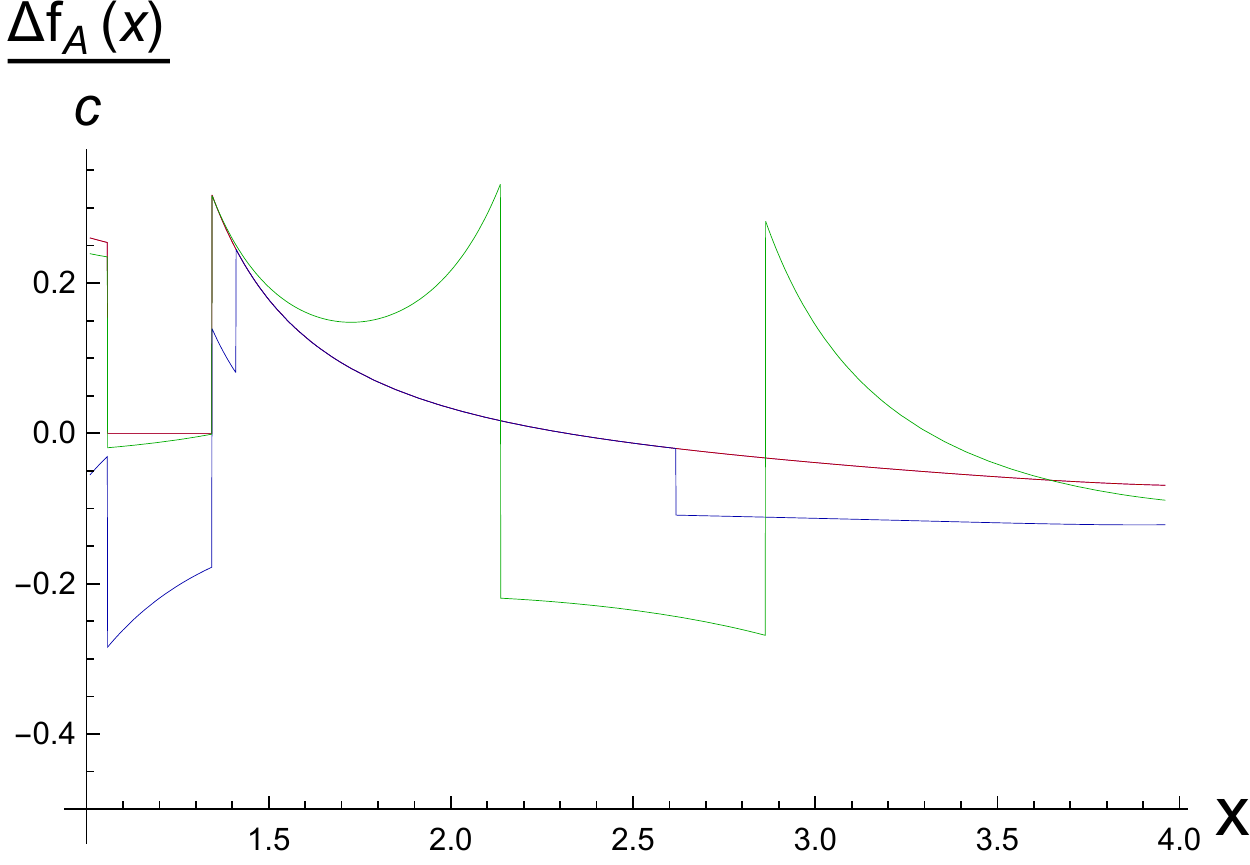}
 \caption{\emph{Left plot}: the evolution of normalized entanglement entropy in the TFD BCFT model minus entropy of the unperturbed ground state. Blue curve corresponds to $S_{bnd}/c=0.16$, the red one to $S_{bnd}/c=0.06$ and magenta to $S_{bnd}/c=0.05$.  \emph{Right plot}: the normalized entanglement contour evolution in the same model and entangling interval fixed to be $x\in (1,4)$ and $S_{bnd}=0$. Curves of different coloring (magenta, red, blue and green) corresponds to different times ($t=-1.5,-0.5,0.5$ and $t=2.5$) respectively.}
 \label{fig:bcftinf}
\end{figure}

In the papers \cite{Rozali:2019day,Sully:2020pza} the pair of 2d BCFT which are in the thermofield double state has been investigated in the context of the Page curve. One can show that this thermofield double state can be represented as a path-integral on the plane where the boundary is situated along the circle ${\cal C}$ in the center. In other words, the thermofield double of 2d BCFT  corresponds to the geometry of the complex plane with the removed disk. It is straightforward to obtain the holographic description of this system and obtain the ETW brane which is the spherical surface hanging from ${\cal C}$ into the bulk. The Lorentzian version of this construction can be obtained by the analytical continuation.
The entanglement entropy evolution in this system can be obtained straightforwardly by conformal mapping $w=w(z)$ to half-plane as in the previous section.
Mapping $w(z)$ of the upper   half-plane on the disk complement of radius $R$ has the form
\be 
w=R\left(\frac{i}{2} + \frac{1}{z+i}\right),\,\,\,\,z=\frac{i}{2}+\frac{R}{i R+w}.
\ee 
The connected and disconnected contributions to the entanglement entropy are given by 
\begin{gather}
S^{cn} =\frac{c}{12} \log \left(\frac{\left|f\left(w_{a}\right)-f\left(w_{b}\right)\right|^{4}}{\epsilon^{4}\left|f^{\prime}\left(w_{a}\right)\right|^{2}\left|f^{\prime}\left(w_{b}\right)\right|^{2}}\right), \\
S^{ds} =\frac{c}{12} \log \left(\frac{16\left(\operatorname{Im} f\left(w_{a}\right)\right)^{2}\left(\operatorname{Im} f\left(w_{b}\right)\right)^{2}}{\epsilon^{4}\left|f^{\prime}\left(w_{a}\right)\right|^{2}\left|f^{\prime}\left(w_{b}\right)\right|^{2}}\right)+2S_b,
\end{gather}
which is a mild generalization of \eqref{eq:EEmirr}.
  We present the entanglement entropy and entanglement contour evolution for BCFT pair entangled in the thermofield in Fig.\ref{fig:bcftinf}. Although the analysis of entanglement entropy behavior in the TFD BCFT model has been performed in \cite{Rozali:2019day} for convenience let us briefly describe its main features here.

The entanglement entropy at some time specified by $R$ for fixed interval size exhibits initial quadratic growth, then it increases linearly and saturates in a non-smooth manner at some time. This seems to be very natural if we assume the thermofield double interpretation of the path-integral geometry. This behavior is very typical in different eternal black holes, Vaidya and other thermalization models  \cite{Ageev:2017wet,Liu:2013iza,Hartman:2013qma}. In contrast to the entanglement entropy late-time decrease in the moving mirror model taking place after the Page time in the TFD BCFT model, it is saturated at some fixed value. In general, the evolution of the entanglement contour resembles the one observed in the mirror model, where numerous `` contour islands'' are spread over the entangling region. 
\section{Concluding remarks and future directions of research}
In this paper, we have shown that the entanglement contour reveals non-trivial fine-grained features of the entanglement entropy in different BCFT setups. We find the presence of spatial entanglement degrees of freedom localization or delocalization in the entanglement entropy due to the presence of a boundary. Referring to the interpretation of BCFT as a model resembling the black hole (as was argued in \cite{Rozali:2019day,Sully:2020pza}) one can conclude that the presence of ``islands in contour'' is an explicit manifestation of ``entanglement island'' phenomena recently discovered in the black hole physics.

It would be interesting to extend this paper in different directions. It would be interesting to calculate the entanglement contour on the CFT side and compare the results for different theories, for example, free against holographic CFTs. The extension of BCFT/entanglement contour setup on the entanglement negativity, Renyi entropy, reflected entropy, and their contour function as well as other related quantum measures also seem to be intriguing. Finally, we would like to clarify the issue related to the local modular flow in BCFTs. The construction in \cite{Wen:2018whg} based on the partial entanglement entropy implies that in some special regions the evolution corresponding to local modular flow should ``freeze'' due to the vanishing of entanglement contour.  Also, it would be interesting to compare the results obtained here with the higher-dimensional BCFT setups \cite{Chu:2021gdb} as well as  the understanding of  entanglement contour role in  the black hole final state paradox (see for example recent interesting proposal in \cite{Arefeva:2021byb} and references therein). 

{\bf Note added:} I became aware of an independent project considering the entanglement contour affected by entanglement islands in the models different from considered here  and also reporting discontinuous and partially vanishing entanglement contour \cite{rolph}.

\section*{Acknowledgements} I would like to thank Andrew Rolph for sharing a draft of his paper \cite{rolph} prior to publication and Qiang Wen for correspondence.


\begin{thebibliography}{} 

\bibitem{Penington:2019npb}
G.~Penington,
``Entanglement Wedge Reconstruction and the Information Paradox,''
JHEP \textbf{09}, 002 (2020)
doi:10.1007/JHEP09(2020)002
[arXiv:1905.08255 [hep-th]].

\bibitem{Almheiri:2019psf}
A.~Almheiri, N.~Engelhardt, D.~Marolf and H.~Maxfield,
``The entropy of bulk quantum fields and the entanglement wedge of an evaporating black hole,''
JHEP \textbf{12}, 063 (2019)
[arXiv:1905.08762 [hep-th]].

\bibitem{Almheiri:2019qdq}
A.~Almheiri, T.~Hartman, J.~Maldacena, E.~Shaghoulian and A.~Tajdini,
``Replica Wormholes and the Entropy of Hawking Radiation,''
JHEP \textbf{05}, 013 (2020)
doi:10.1007/JHEP05(2020)013
[arXiv:1911.12333 [hep-th]].

\bibitem{Almheiri:2020cfm}
A.~Almheiri, T.~Hartman, J.~Maldacena, E.~Shaghoulian and A.~Tajdini,
``The entropy of Hawking radiation,''
[arXiv:2006.06872 [hep-th]].

\bibitem{Chen:2020uac}
H.~Z.~Chen, R.~C.~Myers, D.~Neuenfeld, I.~A.~Reyes and J.~Sandor,
``Quantum Extremal Islands Made Easy, Part I: Entanglement on the Brane,''
JHEP \textbf{10}, 166 (2020)
doi:10.1007/JHEP10(2020)166
[arXiv:2006.04851 [hep-th]].

\bibitem{Chen:2020hmv}
H.~Z.~Chen, R.~C.~Myers, D.~Neuenfeld, I.~A.~Reyes and J.~Sandor,
``Quantum Extremal Islands Made Easy, Part II: Black Holes on the Brane,''
JHEP \textbf{12}, 025 (2020)
doi:10.1007/JHEP12(2020)025
[arXiv:2010.00018 [hep-th]].

\bibitem{Penington:2019kki}
G.~Penington, S.~H.~Shenker, D.~Stanford and Z.~Yang,
``Replica wormholes and the black hole interior,''
[arXiv:1911.11977 [hep-th]].

\bibitem{Almheiri:2012rt}
A.~Almheiri, D.~Marolf, J.~Polchinski and J.~Sully,
``Black Holes: Complementarity or Firewalls?,''
JHEP \textbf{02}, 062 (2013)
doi:10.1007/JHEP02(2013)062
[arXiv:1207.3123 [hep-th]].

\bibitem{Mathur:2005zp}
S.~D.~Mathur,
``The Fuzzball proposal for black holes: An Elementary review,''
Fortsch. Phys. \textbf{53}, 793-827 (2005)
doi:10.1002/prop.200410203
[arXiv:hep-th/0502050 [hep-th]].




\bibitem{Almheiri:2019hni}
A.~Almheiri, R.~Mahajan, J.~Maldacena and Y.~Zhao,
``The Page curve of Hawking radiation from semiclassical geometry,''
JHEP \textbf{03}, 149 (2020)
doi:10.1007/JHEP03(2020)149
[arXiv:1908.10996 [hep-th]].

\bibitem{Page:1993wv}
D.~N.~Page,
``Information in black hole radiation,''
Phys. Rev. Lett. \textbf{71}, 3743-3746 (1993)
doi:10.1103/PhysRevLett.71.3743
[arXiv:hep-th/9306083 [hep-th]].

\bibitem{Page:1993df}
D.~N.~Page,
``Average entropy of a subsystem,''
Phys. Rev. Lett. \textbf{71}, 1291-1294 (1993)
doi:10.1103/PhysRevLett.71.1291
[arXiv:gr-qc/9305007 [gr-qc]].





\bibitem{Ryu:2006bv}
 S.~Ryu and T.~Takayanagi,
  ``Holographic derivation of entanglement entropy from AdS/CFT,''
  Phys.\ Rev.\ Lett.\  {\bf 96}, 181602 (2006)
  [hep-th/0603001].

\bibitem{Hubeny:2007xt} 
  V.~E.~Hubeny, M.~Rangamani and T.~Takayanagi,
  ``A Covariant holographic entanglement entropy proposal,''
  JHEP {\bf 0707}, 062 (2007)
  doi:10.1088/1126-6708/2007/07/062
  [arXiv:0705.0016 [hep-th]].
  
\bibitem{Yoshida:2019qqw}
B.~Yoshida,
``Firewalls vs. Scrambling,''
JHEP \textbf{10}, 132 (2019)
[arXiv:1902.09763 [hep-th]].

\bibitem{chen-vidal}
Y.~{Chen} and G.~{Vidal}, ``{Entanglement contour},'' { Journal of
  Statistical Mechanics: Theory and Experiment}, vol.~10, p.~10011, Oct. 2014, arXiv:
  

\bibitem{Tonni:2017jom} 
  E.~Tonni, J.~Rodríguez-Laguna and G.~Sierra,
  ``Entanglement hamiltonian and entanglement contour in inhomogeneous 1D critical systems,''
  J.\ Stat.\ Mech.\  {\bf 1804}, no. 4, 043105 (2018)
  doi:10.1088/1742-5468/aab67d
  [arXiv:1712.03557 [cond-mat.stat-mech]].

\bibitem{Coser:2017dtb} 
  A.~Coser, C.~De Nobili and E.~Tonni,
  ``A contour for the entanglement entropies in harmonic lattices,''
  J.\ Phys.\ A {\bf 50}, no. 31, 314001 (2017)
  doi:10.1088/1751-8121/aa7902
  [arXiv:1701.08427 [cond-mat.stat-mech]].

\bibitem{Eisler:2019rnr} 
  V.~Eisler, E.~Tonni and I.~Peschel,
  ``On the continuum limit of the entanglement Hamiltonian,''
  J.\ Stat.\ Mech.\  {\bf 1907}, no. 7, 073101 (2019)
  doi:10.1088/1742-5468/ab1f0e
  [arXiv:1902.04474 [cond-mat.stat-mech]].

\bibitem{Wen:2018whg} 
  Q.~Wen,
  ``Fine structure in holographic entanglement and entanglement contour,''
  Phys.\ Rev.\ D {\bf 98}, no. 10, 106004 (2018)
  doi:10.1103/PhysRevD.98.106004
  [arXiv:1803.05552 [hep-th]].

\bibitem{Wen:2019ubu} 
  Q.~Wen,
  ``Entanglement contour from subset entanglement entropies,''
  arXiv:1902.06905 [hep-th].

\bibitem{Kudler-Flam:2019oru} 
  J.~Kudler-Flam, I.~MacCormack and S.~Ryu,
  ``Holographic entanglement contour, bit threads, and the entanglement tsunami,''
  arXiv:1902.04654 [hep-th].

\bibitem{Ageev:2021iyw}
D.~S.~Ageev,
``On the entanglement contour of excited states in the holographic CFT,''
Eur. Phys. J. Plus \textbf{136}, no.4, 435 (2021)
doi:10.1140/epjp/s13360-021-01397-w
[arXiv:1905.06920 [hep-th]].

\bibitem{Han:2019scu} 
  M.~Han and Q.~Wen,
  ``Entanglement entropies from entanglement contour: annuli and spherical shells,''
  arXiv:1905.05522 [hep-th].

\bibitem{Han:2021ycp}
M.~Han and Q.~Wen,
``First Law and Quantum Correction for Holographic Entanglement Contour,''
[arXiv:2106.12397 [hep-th]].

\bibitem{MacCormack:2020auw}
I.~MacCormack, M.~T.~Tan, J.~Kudler-Flam and S.~Ryu,
``Operator and entanglement growth in non-thermalizing systems: many-body localization and the random singlet phase,''
[arXiv:2001.08222 [cond-mat.str-el]].

\bibitem{Kudler-Flam:2019nhr} 
  J.~Kudler-Flam, H.~Shapourian and S.~Ryu,
  ``The negativity contour: a quasi-local measure of entanglement for mixed states,''
  arXiv:1908.07540 [hep-th].
  
  

\bibitem{cardy}
Cardy, J. L. (1984). Conformal invariance and surface critical behavior. Nuclear Physics B, 240(4), 514-532.

\bibitem{Rozali:2019day}
M.~Rozali, J.~Sully, M.~Van Raamsdonk, C.~Waddell and D.~Wakeham,
``Information radiation in BCFT models of black holes,''
JHEP \textbf{05}, 004 (2020)
[arXiv:1910.12836 [hep-th]].

\bibitem{Sully:2020pza}
J.~Sully, M.~V.~Raamsdonk and D.~Wakeham,
``BCFT entanglement entropy at large central charge and the black hole interior,''
JHEP \textbf{03}, 167 (2021)
[arXiv:2004.13088 [hep-th]].

\bibitem{Akal:2020twv}
I.~Akal, Y.~Kusuki, N.~Shiba, T.~Takayanagi and Z.~Wei,
``Entanglement Entropy in a Holographic Moving Mirror and the Page Curve,''
Phys. Rev. Lett. \textbf{126}, no.6, 061604 (2021)
doi:10.1103/PhysRevLett.126.061604
[arXiv:2011.12005 [hep-th]].

\bibitem{Akal:2021foz}
I.~Akal, Y.~Kusuki, N.~Shiba, T.~Takayanagi and Z.~Wei,
``Holographic moving mirrors,''
[arXiv:2106.11179 [hep-th]].

\bibitem{BD}
Birrell, N. D., Birrell, N. D.,  Davies, P. C. W. (1984). Quantum fields in curved space.

\bibitem{mir1}
Fulling, S. A.,  Davies, P. C. (1976). Radiation from a moving mirror in two dimensional space-time: conformal anomaly. Proceedings of the Royal Society of London. A. Mathematical and Physical Sciences, 348(1654), 393-414.

\bibitem{mir2}
M.~R.~R.~Good, E.~V.~Linder and F.~Wilczek,
``Moving mirror model for quasithermal radiation fields,''
Phys. Rev. D \textbf{101}, no.2, 025012 (2020)
[arXiv:1909.01129 [gr-qc]].

\bibitem{Bianchi:2014qua}
E.~Bianchi and M.~Smerlak,
``Entanglement entropy and negative energy in two dimensions,''
Phys. Rev. D \textbf{90}, no.4, 041904 (2014)
doi:10.1103/PhysRevD.90.041904
[arXiv:1404.0602 [gr-qc]].

\bibitem{Hotta:2015huj}
M.~Hotta and A.~Sugita,
``The Fall of Black Hole Firewall: Natural Nonmaximal Entanglement for Page Curve,''
PTEP \textbf{2015}, no.12, 123B04 (2015)
[arXiv:1505.05870 [gr-qc]].

\bibitem{Good:2016atu}
M.~R.~R.~Good, K.~Yelshibekov and Y.~C.~Ong,
``On Horizonless Temperature with an Accelerating Mirror,''
JHEP \textbf{03}, 013 (2017)
[arXiv:1611.00809 [gr-qc]].

\bibitem{Chen:2017lum}
P.~Chen and D.~h.~Yeom,
``Entropy evolution of moving mirrors and the information loss problem,''
Phys. Rev. D \textbf{96}, no.2, 025016 (2017)
[arXiv:1704.08613 [hep-th]].

\bibitem{Takayanagi:2011zk}
T.~Takayanagi,
``Holographic Dual of BCFT,''
Phys. Rev. Lett. \textbf{107}, 101602 (2011)
[arXiv:1105.5165 [hep-th]].

\bibitem{Fujita:2011fp}
M.~Fujita, T.~Takayanagi and E.~Tonni,
``Aspects of AdS/BCFT,''
JHEP \textbf{11}, 043 (2011)
[arXiv:1108.5152 [hep-th]].


\bibitem{Liu:2013iza}
H.~Liu and S.~J.~Suh,
``Entanglement Tsunami: Universal Scaling in Holographic Thermalization,''
Phys. Rev. Lett. \textbf{112}, 011601 (2014)
[arXiv:1305.7244 [hep-th]].

\bibitem{Hartman:2013qma}
T.~Hartman and J.~Maldacena,
``Time Evolution of Entanglement Entropy from Black Hole Interiors,''
JHEP \textbf{05}, 014 (2013)
[arXiv:1303.1080 [hep-th]].

\bibitem{Ageev:2017wet}
D.~S.~Ageev and I.~Y.~Aref'eva,
``Holographic Non-equilibrium Heating,''
JHEP \textbf{03}, 103 (2018)
[arXiv:1704.07747 [hep-th]].

\bibitem{Chu:2021gdb}
J.~Chu, F.~Deng and Y.~Zhou,
``Page Curve from Defect Extremal Surface and Island in Higher Dimensions,''
[arXiv:2105.09106 [hep-th]].

\bibitem{Arefeva:2021byb}
I.~Aref'eva and I.~Volovich,
``Quantum explosions of black holes and thermal coordinates,''
[arXiv:2104.12724 [hep-th]].

\bibitem{rolph}
Andrew Rolph, TBA

\end{thebibliography}
\end{document}